\documentclass[%
  aps,            
  prx,            
  reprint,        
  superscriptaddress, 
  amsmath,amssymb, 
  longbibliography,
]{revtex4-2}

\usepackage{graphicx}   
\usepackage{dcolumn}    
\usepackage{bm}         
\usepackage[colorlinks=true, allcolors=blue]{hyperref}
\usepackage{braket}
\usepackage{booktabs}

\usepackage{siunitx}
\usepackage[T1]{fontenc}

\NewDocumentCommand{\figref}{m o}{%
  \IfNoValueTF{#2}
    {\hyperref[#1]{\ref*{#1}}}
    {\hyperref[#1]{\ref*{#1}#2}}%
}
 
\begin{document}

\title{Programmable cavity QED with a fiber-integrated atomic array}

\newcommand{\ista}{\affiliation{Institute of Science and Technology Austria (ISTA), Am Campus 1, 3400 Klosterneuburg, Austria}}
\newcommand{\tuwien}{\affiliation{Vienna Center for Quantum Science and Technology, Atominstitut, TU Wien, Vienna, Austria}}

\author{Stephan~Roschinski}
\thanks{These authors contributed equally to this work.}\ista\tuwien
\author{Johannes~Schabbauer}
\thanks{These authors contributed equally to this work.}\ista\tuwien
\author{Franz~von~Silva-Tarouca}\ista\tuwien
\author{Marvin~Holten}\tuwien
\author{Damien~Bloch}\ista\tuwien
\author{Julian~L\'eonard}\email{julian.leonard@ista.ac.at} \ista\tuwien

\begin{abstract}
Strong atom-photon interactions in optical cavities are a key resource for quantum information processing, quantum networking, and the exploration of quantum optical effects. Optical tweezer arrays offer scalable, site-resolved control of neutral atoms, but their integration with high-cooperativity cavity QED systems remains challenging. Here we combine a twelve-site $^{87}$Rb optical tweezer array with a high-cooperativity fiber Fabry-Pérot microcavity. The array is positioned within the cavity mode and individual sites are controlled with subwavelength precision, enabling continuous tuning of the single-atom coupling strength via deterministic displacement through the standing-wave field. For up to five atoms coupled to the cavity, we measure collectively enhanced vacuum Rabi splitting and implement cavity-based non-destructive readout of the number of coupled atoms. These results establish a scalable architecture for cavity-mediated entanglement generation and many-body cavity QED with single-atom control, and they lay the foundation for fiber-integrated quantum network nodes.
\end{abstract}

\maketitle

\section{Introduction}
Linking quantum processors into a single network is a fascinating approach to overcome the scalability limits of independent devices \cite{Kimble2008, Wehner2018}, thereby paving the way for distributed quantum computation \cite{Jiang2007, Monroe2014}, secure communication \cite{Pirandola2020}, and network-enhanced sensing \cite{Gottesman2012, Komar2014}. In such a network, photons act as flying qubits, carrying quantum states and entanglement between nodes over fiber-optic interconnects. The nodes contain matter qubits that couple coherently to individual photons, converting quantum information between light and matter with high fidelity. Neutral atoms coupled to optical cavities have emerged as a powerful platform for this vision, as they offer both precise control of atomic qubits and efficient atom-photon interfaces \cite{reisererCavitybasedQuantumNetworks2015}. 

Beyond providing photonic connectivity, each node must generate, store, and process quantum information locally \cite{Covey2023}. Optical tweezer arrays enable the assembly \cite{Barredo2016, Endres2016} and quantum control \cite{Tsai2025, Senoo2026, evered2026highfidelityentanglinggatesnonlocal, lin2026sustaininghighfidelityquantumlogic} of defect-free registers of individually addressable neutral atoms. Coupling these atomic tweezer arrays with a common cavity mode therefore provides a route toward quantum network nodes with the desired local processing power.

The central challenge in combining optical cavities with tweezer arrays is geometric. Optical access required for tweezer focusing, site-resolved control, and atom imaging conflicts with the small cavity mode diameters needed for high single-atom cooperativity. A common strategy is to operate a macroscopic Fabry-Pérot resonator near the concentric stability limit, where a large mirror separation preserves optical access while still yielding a small mode waist. Integration of such macroscopic cavities with individually controllable atom arrays has only recently been realized experimentally \cite{yanSuperradiantSubradiantCavity2023a, Liu2023, hartungQuantumnetworkRegisterAssembled2024, Zhang2024CavityDarkMode, Hu2025, Peters2025, santisRealizationCavitycoupledRydberg2026, Wang2025}. 

Significantly larger cooperativities can be achieved by miniaturizing the cavity and further reducing the mode waist \cite{Hunger2010, samutpraphoot2020, Ruddell20}. Reaching these high cooperativities is paramount, as they determine the efficiency of the atom-photon interface. Moreover, the resulting small mode waists become comparable to optical fiber mode diameters, naturally enabling fiber-integrated implementations. These advantages, however, come at the cost of proximity to the cavity structure or restricted optical access. Consequently, experiments in such systems have largely been limited to one \cite{Kato2015, gallegoStrongPurcellEffect2018, Brekenfeld2020, wangUltrafastHighFidelityState2025} or two trapped atoms \cite{Dordevic2021, grinkemeyerErrorDetectedQuantumOperations2024}. Uniting the control in atomic tweezer arrays with the high cooperativity of microscopic cavities has remained elusive.

\begin{figure*}
    \includegraphics{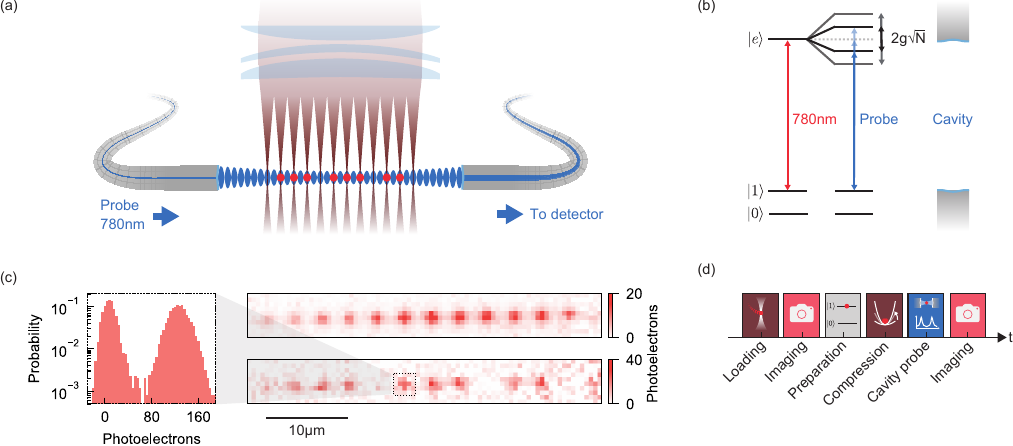}
    \caption{Realization of a fiber-integrated atomic array. (a)~Schematic of the experimental setup. Two mirrors, fabricated on the end-facets of optical fibers, form a high-finesse optical cavity. Optical tweezers are projected through a microscope objective onto the cavity mode and loaded with $^{87}\mathrm{Rb}$ atoms. A resonant probe beam is coupled into the cavity through one fiber mirror, and the transmitted light is detected on a single-photon counting module through the other.
    (b)~Relevant level scheme for the $D_2$-line of $^{87}\mathrm{Rb}$, with qubit states $\ket{0}=\ket{F=1, m_F=1}$, $\ket{1}=\ket{F=2, m_F=2}$ and excited state $\ket{e}=\ket{F'=3, m_{F'}=3}$. For all measurements, irrespective of atom number $N$, the cavity is resonant with the atomic transition from $\ket{1}$ to $\ket{e}$. The single-atom coupling strength $g$ between the cavity mode and the $\ket{1}\rightarrow \ket{e}$ transition gives rise to a vacuum Rabi splitting of $2g \sqrt{N}$.
    (c)~Fluorescence imaging. Averaged (top) and single shot (bottom) fluorescence picture of the 12-site atom array. A typical histogram of a single-site signal shows high-fidelity discrimination between singly occupied and empty sites. We extract an average imaging fidelity of $\SI{99.91(3)}{\percent}$ by fitting the bimodal distribution. 
    (d)~Typical experimental sequence. We first load the tweezers and acquire a fluorescence image to identify occupied traps. After optical pumping, we ramp up the optical power in each tweezer to reduce the atomic wavefunction spread, and then probe the cavity transmission. Finally, we take a second fluorescence image to verify that the traps remain occupied.}
    \label{fig1}
\end{figure*}

Here, we realize a platform that achieves this goal and combines an atomic tweezer array with an optical microcavity with high cooperativity. The cavity is built from fiber mirrors, featuring a native fiber-coupled quantum interface between individually controlled atoms and propagating photons. We demonstrate key capabilities of this system, including programmable single-atom coupling, collectively enhanced atom-photon interactions, and nondestructive atom-number measurements, enabling future applications in quantum networking and many-body cavity QED with individually controllable atoms.

\section{A fiber-integrated atomic array}

The experimental platform consists of a fiber Fabry-Pérot cavity integrated with a one-dimensional optical tweezer array of $^{87}$Rb atoms (Fig.~\figref{fig1}[a]). It builds on previous experiments with fiber-based Fabry-Pérot cavities coupled to atomic ensembles \cite{Colombe2007}, single atoms \cite{gallegoStrongPurcellEffect2018, Brekenfeld2020, wangUltrafastHighFidelityState2025}, and two atoms \cite{grinkemeyerErrorDetectedQuantumOperations2024}, and extends them to a multi-site tweezer-array setting. The cavity has a length of $\SI{85}{\micro \meter}$ and consists of two fiber mirrors with radii of curvature of \SI{200}{\micro \meter}. Its finesse $\mathcal{F} = 29750$ and $1/e^2$ mode waist $w_0 = \SI{4.5}{\micro \meter}$ (Appendix~\ref{app:cavity-characterization}) yield a geometric single-atom cooperativity $\eta = 24\mathcal{F}/(\pi k^2 w_0^2)=173$. Here $k=2\pi/\lambda$ is the wavenumber at the cavity wavelength $\lambda=\SI{780}{\nano \meter}$, resonant with the $^{87}$Rb transition $\ket{1} = \ket{F=2, m_F = 2} \rightarrow \ket{e} = \ket{F'=3, m_{F'}=3}$ (Fig.~\figref{fig1}[b]). One mirror is fabricated on the end facet of a single-mode fiber and is used to send a probe pulse into the cavity, the other on the end facet of a multimode fiber to enhance the photon outcoupling rate. We detect the photons leaving the cavity through the outcoupling mirror on a single-photon counting module (SPCM). From spectral measurements (Fig.~\figref{fig_g_control}) we determine the single atom cavity QED parameters $(g, \kappa, \gamma) =  2 \pi \times (107(4), 29.6(2), 3.03)\,\si{\mega \hertz}$, where $g$ is the maximum single atom-cavity coupling strength, and $\kappa$ ($\gamma$) is the cavity field (atomic coherence) decay rate. These parameters place our system deep in the strong-coupling regime. The measured coupling strength is comparable to the maximum expected value of $g_0=2\pi\times \SI{124.7}{\mega \hertz}$, with the reduction attributable to the finite spatial extent of the atomic wavepacket, and possible imperfections in atomic state preparation and polarization of the probe light.

The cavity mode lies in the focal plane of a microscope objective with numerical aperture $\mathrm{NA}=0.5$, which provides optical access for single-atom resolved trapping and fluorescence readout. We use this objective to project a one-dimensional array of up to twelve optical tweezers onto the cavity mode. The tweezers are generated from a laser beam at $\SI{810}{\nano \meter}$ with an acousto-optic deflector (AOD), whose radio-frequency tones independently set the position of each trap along the cavity axis. This site-resolved positioning allows atoms to be placed at selected phases of the standing-wave cavity field and thereby controls the local atom-cavity coupling. We load the tweezers stochastically from a cloud of laser-cooled $^{87}\mathrm{Rb}$ atoms and optically pump them in state $\ket{1}$.

Site-resolved fluorescence imaging is essential for tweezer-array assembly and readout, but it becomes challenging near compact photonic devices because imaging light can scatter from nearby surfaces, requiring specialized imaging methods \cite{Kim2019, Meng2020, Menon2024}. In our fiber-cavity geometry, the mirror separation is large enough for a pair of counter-propagating imaging beams to pass without clipping on the mirrors, enabling direct, single-shot fluorescence imaging of the full atomic array. The high-NA microscope objective collects the atomic fluorescence, which is imaged onto a low-noise camera (Fig.~\figref{fig1}[c]). From these images, we obtain a mean single-site loading probability of $\SI{56}{\percent}$. A histogram of the site-resolved photon counts shows a bimodal distribution, from which we extract an average imaging fidelity of $\SI{99.91(3)}{\percent}$ for discriminating empty and occupied tweezers (Appendix~\ref{app:imaging}).

\begin{figure}
    \centering
    \includegraphics{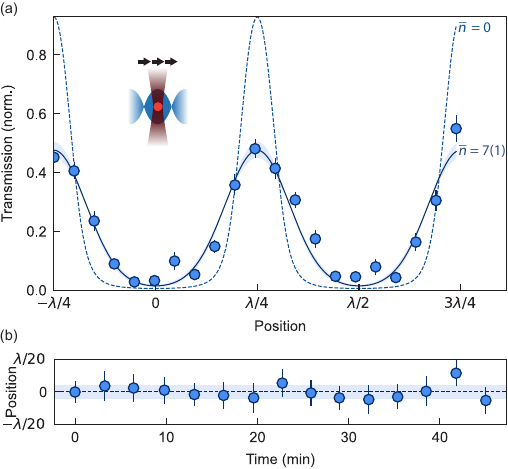}
    \caption{Position-controlled cavity blockade. (a)~Cavity transmission as the tweezer is translated along the cavity axis. The transmission reflects the local atom-photon coupling: atoms near field nodes leave the cavity highly transmissive, whereas atoms near antinodes suppress transmission through enhanced coupling to the cavity mode. The data agree with an exact finite-temperature theory, taking into account atomic zero-point motion and a mean motional occupation in the trap of $\bar{n}=7(1)$ (solid line). (b)~Repeated measurements of the atomic position along the cavity axis show a variation of $1.1\%\times\lambda$, well below the spatial period $\lambda/2$ of the standing-wave mode.
    }
    \label{fig:transmission}
\end{figure}

\section{Position-controlled cavity blockade}
The position of an atom within the cavity standing wave directly determines its coupling strength and therefore its nonlinear optical response. We exploit this dependence to realize a position-controlled cavity blockade using a single tweezer-trapped atom (Fig.~\figref{fig:transmission}). A single-atom nonlinearity is a hallmark of cavity QED, in which strong atom-photon coupling renders the optical response of a cavity nonlinear at the level of individual photons. This mechanism also forms the basis for many quantum gates \cite{reisererCavitybasedQuantumNetworks2015}. For an atom resonant with the cavity, the atom-photon interaction hybridizes the bare cavity and atomic excitations into two dressed modes, suppressing transmission at the empty-cavity resonance. In the weak-probe limit, this suppression is governed by the local single-atom cooperativity $\eta(x)=g(x)^2/\kappa\gamma$, where $g(x)=g_0 \cos(kx)$ is the position-dependent atom-cavity coupling and $x$ is the position of the atom relative to a cavity antinode. For resonant probing, the transmitted intensity normalized by the empty cavity transmission is $T(x) = [1+\eta (x)]^{-2}$. The cooperativity $\eta(x)=\eta_0\cos^2(kx)$ is therefore modulated with period $\lambda/2$: atoms at antinodes couple maximally and  suppress the resonant transmission, whereas atoms placed near nodes have minimal coupling and approach the empty-cavity response. In this way, the blockade response maps the local cavity cooperativity sampled by the atom.

We probe this position-dependent blockade by translating a single trapped atom along the cavity axis while measuring the transmission of a near-resonant probe pulse (Fig.~\figref{fig:transmission}[a]). The transmission oscillates with the expected standing-wave period of $\lambda/2$, reaching a minimum at antinodes, where the cooperativity is maximal, and approaching the empty-cavity response near nodes. The modulation contrast is reduced below the expected suppression, which is quantitatively explained by the finite spatial extent of the atomic wavepacket from zero-point and thermal motion
(Appendices~\ref{app:tweezer-characterization}, \ref{app:wavefunction-spread}). The resulting model reproduces the data across the full scan without free parameters, confirming programmable control of the single-atom cooperativity. 

Programming the cavity blockade with a steerable tweezer requires subwavelength stability between the tweezer position and the cavity standing wave \cite{seubertTweezerAssistedSubwavelengthPositioning2025}. Such stability is also essential for multi-atom schemes in which the relative optical phase between atoms determines the collective response, such as dark states and super- or subradiant modes. Often this stability is achieved by pinning the atoms with an intracavity lattice, however, this restricts the atom-cavity coupling to discrete values. In our experiment, the atoms remain confined solely by the optical tweezer, enabling continuous control of the atom-cavity coupling. We quantify the relative stability by repeatedly measuring the position-dependent blockade and extracting the global phase of the transmission modulation (Fig.~\figref{fig:transmission}[b]). Over the measurement time, the inferred relative displacement has a standard deviation of $1.1\% \times \lambda$, well below both the standing-wave period and the wavepacket size. This stability establishes optical tweezers as a viable alternative to intracavity lattices for precision control of
atom-cavity coupling.

\begin{figure}
    \centering
    \includegraphics{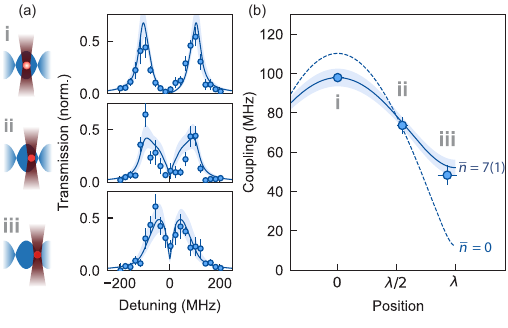}
    \caption{Programmable atom-photon coupling. (a) Cavity transmission spectra for three representative atomic positions along the standing-wave mode. By translating a single atom through the cavity field, we continuously tune the coupling from the strong-coupling regime at an antinode (i), where a clear vacuum Rabi splitting appears, through intermediate coupling between node and antinode (ii), to weak coupling near a node (iii), where the spectrum approaches that of an empty cavity. Solid lines are fits with the coupling strength as the only free parameter.
    (b) Extracted coupling strength versus position. The data agree with an exact finite-temperature theory (solid lines).}
    \label{fig_g_control}
\end{figure}

\section{Programmable coupling}
The stable tweezer-cavity positioning enables direct control of the coherent atom-photon coupling of a single trapped atom (Fig.~\figref{fig_g_control}). Whereas resonant transmission probes the local cooperativity, cavity spectroscopy measures the coupling strength $g(x)$ itself through the vacuum Rabi splitting. In the strong-coupling regime, the atom and cavity hybridize into two polariton modes separated by $2g(x)$. Figure~\figref{fig_g_control}[a] shows transmission spectra for three representative atomic positions. As the atom is translated from an antinode toward a node of the standing-wave field, the vacuum Rabi splitting decreases continuously and the spectrum evolves toward that of an empty cavity. The residual splitting near the node is quantitatively explained by the finite spatial extent of the atomic wavepacket (Appendix~\ref{app:wavefunction-spread}), which samples regions of nonzero cavity field.

We extract the local coupling strength by fitting the transmission spectra using the previous model (Fig.~\figref{fig_g_control}[b]). The extracted $g$ follows the expected standing-wave dependence as the tweezer is translated. The tuning range extends from maximal coupling at an antinode to a residual coupling near a node, with the lower bound set by the finite spatial extent of the atomic wavepacket. Tighter confinement or lower temperature would reduce this motional averaging and extend the dynamic range toward zero coupling. We also investigate the stability of the maximal coupling strength, where many cavity-QED protocols operate, and observe relative variations of $2.4\%$ over about twelve hours (Appendix~\ref{app:stability}). These measurements demonstrate direct, robust control of the atom-cavity coupling through the tweezer position. 

\begin{figure}
    \centering
    \includegraphics{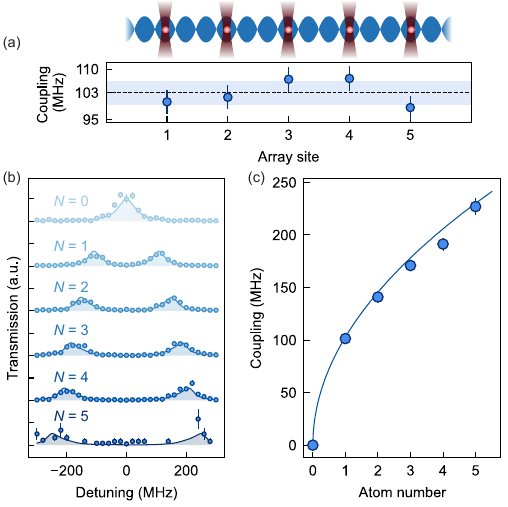}
    \caption{Collective enhancement of the coupling. (a)~We project a five-site tweezer array with all sites positioned near antinodes and measure the coupling strength of each site individually. The extracted couplings agree within error, establishing the homogeneity of the coupling control. (b)~Cavity transmission spectra for atom numbers $N=0-5$. As additional  tweezers are occupied, the vacuum Rabi splitting increases due to collective enhancement of the atom-cavity coupling. Solid lines are fits to the data with the coupling strength as the only free parameter. (c)~Collective coupling strength extracted from the spectra in (b). The solid line shows the expected Tavis-Cummings scaling, $g_\text{eff}=\sqrt{N}g$, using the average single-atom coupling $g$ from (a).}
    \label{fig3}
\end{figure}

\section{Collective coupling}
The site-resolved control of individual atom-cavity couplings naturally extends to collective interactions in atomic arrays. When several atoms couple to the same cavity mode, the optical response is governed by a collective excitation of the array. For atoms with local coupling strengths $g_i$, the Tavis-Cummings model predicts an effective coupling $g_\text{eff} = \sqrt{\sum_ig_i^2}$. In the homogeneous limit, when each atom couples maximally, this gives the characteristic collective enhancement $g_\text{eff} = g\sqrt{N}$ and a vacuum Rabi splitting $2g_\text{eff}$ \cite{Tavis1968, Blaha2022}. Positioning individual tweezers within the cavity standing wave turns the collective coupling of an atomic register into a programmable quantity set by the site-resolved atom-photon couplings. This control can also compensate slow spatial variations of the cavity-mode amplitude, for example those arising from the gaussian mode envelope or divergence of the mode. This programmable setting contrasts with Tavis-Cummings realizations in ensemble cavity QED, where the atoms sample the spatially varying cavity coupling according to their trapping distribution \cite{Brennecke2007, Colombe2007}.

To realize this regime experimentally, we first project a five-site tweezer array in which each site is positioned near an antinode of the cavity standing wave. We independently measure each single-atom coupling from the transmission spectrum and find a mean coupling of $\SI{103(4)}{\mega \hertz}$, with variations below $\SI{4}{\percent}$ across the sites (Fig.~\figref{fig3}[a]). This confirms homogeneous coupling across the array. We then load the five-site array and record cavity transmission spectra (Fig.~\figref{fig3}[b]). As the atom number increases, the spectrum develops a vacuum Rabi doublet with increasing splitting, signaling the collective enhancement of the coupling. We quantify the enhancement by fitting the transmission spectra with our cavity-QED model and extract $g_\text{eff}$ for each atom number. The extracted collective couplings follow the characteristic $\sqrt{N}$ scaling of the Tavis-Cummings model (Fig.~\figref{fig3}[c]), demonstrating that the programmed site-resolved couplings determine the collective strong-coupling response of the array.

\begin{figure}
    \centering
    \includegraphics{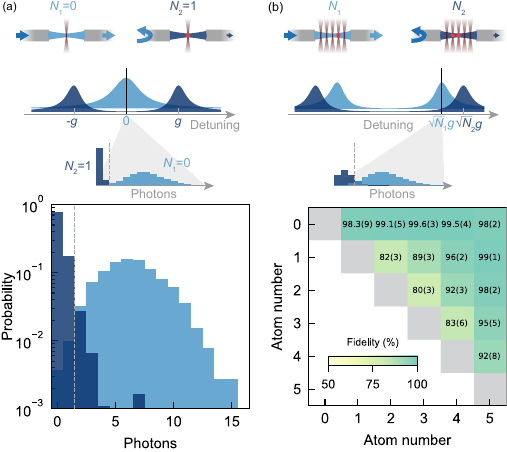}
    \caption{Non-destructive atom-number readout. (a) We detect the atom number from the suppression of resonant probe light due to cavity blockade: for a single atom, $N=1$, the transmission is reduced (dark blue), whereas the uncoupled case, $N=0$, remains highly transmissive (light blue). The dashed line shows the photon number for optimal threshold. 
    (b) While resonant probe light is reflected for any atom number $N\geq 1$, probe light that is resonant with the polariton peak for $N$ atoms is sensitive to the respective atom number. Measuring the cavity then at this frequency we extract histograms as in (a).
    Using this procedure we extract fidelities for every pair of atom numbers. Combinations with greater frequency difference between their respective polariton peaks yield the highest fidelities.}
    \label{fig5}
\end{figure}

\section{Atom number readout}
The collective coupling of a homogeneous atomic array provides a direct spectroscopic handle on the number of coupled atoms. We exploit this dependence for cavity-assisted atom-number detection. Fast, non-destructive measurements are a key capability for quantum information processing, enabling mid-circuit readout, feedback, and error detection without removing atoms from the register. Related cavity-based readout techniques have been demonstrated for single atoms using resonant light \cite{Bochmann2010, Gehr2010}, and for multiple atoms in the dispersive regime \cite{McKeever2004, Peters2025}. Because the collective coupling scales as $g_\text{eff}= g\sqrt{N}$ for nearly homogeneous site couplings, different atom numbers give rise to distinct collective vacuum Rabi splittings. A probe at a suitable detuning can therefore discriminate between atom numbers through the cavity transmission. In an occupied register, the same measurement can be interpreted as a state-resolved population measurement: atoms in $\ket{1}$ couple resonantly to the cavity, whereas atoms in $\ket{0}$ are far detuned and effectively dark. Thus, conditioned on the known tweezer occupancy, the inferred atom number corresponds to the number of atoms in the coupled state $\ket{1}$. Such collective, nondestructive measurements of the excitation number form the basis of many cavity-mediated entanglement protocols, where projection onto a subspace can generate entanglement between atoms \cite{haasEntangledStatesMore2014, Chen2015, WelteCarving2017}.

We first illustrate the readout principle by distinguishing an empty array from a single coupled atom (Fig.~\figref{fig5}[a]). We prepare a coupled tweezer array with either $N=0$ or $N=1$ and probe the cavity at the empty-cavity resonance while recording the transmitted photons. For $N=0$, the cavity is resonant with the probe and transmits strongly. For $N=1$, the strong coupling suppresses resonant transmission, producing a markedly lower photon count. A threshold on the detected photon count discriminates between zero and one atom with fidelity $97.4(6)\%$ in $\SI{40}{\micro \second}$ and $99.0(4)\%$ in $\SI{100}{\micro \second}$. 

We then extend this protocol to pairwise discrimination of larger atom numbers (Fig.~\figref{fig5}[b]). For a chosen pair $N_1$ and $N_2$, we set the probe detuning on resonance to the collective vacuum Rabi resonance for one atom number, for instance $\delta= g\sqrt{N_1}$ for the upper polariton of $N_1$. This results in high transmission for atom number $N_1$ and low transmission for atom number $N_2$. Repeating the experiment yields photon-count distributions from which we extract the atom-number readout fidelity from the overlap of the Poisson distributions (Appendix~\ref{app:bayesian-analysis}). The observed performance is consistent with the finite overlap expected from photon shot noise, indicating that higher detected photon numbers would further improve the discrimination fidelity.

This resonant readout is complementary to the dispersive measurements commonly used in cavity QED. In the dispersive regime, atom number is inferred from an atom-induced cavity shift, which for homogeneous coupling scales as $Ng^2/\Delta_\text{a}$. Here, we instead operate in the resonant strong-coupling regime and use the collective polariton resonances as the readout basis. Since these resonances occur at $\delta \simeq \pm g\sqrt{N}$, two atom numbers $N_1$ and $N_2$ are separated by $(\sqrt{N_2}-\sqrt{N_1})g$. When this separation exceeds the corresponding dispersive shift, probing near a collective resonance converts atom number into a larger transmission contrast. This makes polariton-resolved readout particularly natural for high-cooperativity microscopic cavities, where the normal modes are well resolved.

\section{Outlook}
We have integrated a high-cooperativity fiber Fabry-Pérot cavity with an optical tweezer array and demonstrated site-resolved control of atoms within the cavity mode. Several direct improvements can extend the capabilities demonstrated here. Deterministic rearrangement enables defect-free arrays to be prepared inside the cavity mode, while slightly longer cavities increase the number of atoms that can be coupled to the resonator. Improved motional control, through lower temperatures or tighter confinement, reduces wavepacket averaging and increases the dynamic range over which individual couplings can be tuned. Higher photon collection efficiency and optimized cavity outcoupling further improve the speed and fidelity of cavity-assisted readout \cite{wangUltrafastHighFidelityState2025}. In addition to moving atoms through the standing wave, site-selective AC Stark shifts could hide selected atoms from the cavity by shifting their optical transition out of resonance, enabling subset-selective readout and coupling without physically rearranging the array \cite{Hu2025, lee2026rapidcavitybasedmidcircuitmeasurement}.

The ability to set the atom-cavity coupling site by site opens the door to engineered collective atom-photon interfaces with broad applications. Homogeneous coupling produces an atom-number-dependent polariton spectrum for nondestructive readout, whereas deliberately inhomogeneous coupling patterns could address selected subsets of atoms, engineer tailored collective modes, or realize weighted spin models and programmable cavity-mediated interactions \cite{Periwal2021, Ye2023, Shapira2025, Lu2025}. The ability to program the atom-light coupling also enables non-local cavity-mediated quantum gates \cite{Borregaard2015, Chen2015, Ramette2025}. Combining this control with Rydberg excitations \cite{santisRealizationCavitycoupledRydberg2026} would create a hybrid architecture in which local Rydberg gates are complemented by non-local cavity-mediated gates \cite{Ramette2022}. Such capabilities are relevant for measurement-based quantum computing \cite{Raussendorf2001, Briegel2009}, dissipative and measurement-induced entanglement generation \cite{Kastoryano2011, haasEntangledStatesMore2014}, quantum error correction \cite{Reiter2017}, and the preparation or distillation of scalable entangled states \cite{Zhao2021, Thomas2022, Jandura2024, Nagib2025}. They also provide a platform for studying cavity-mediated many-body physics and nonequilibrium dynamics with microscopic control over the atomic positions and couplings \cite{Zhang2013, Gelhausen2016, Rohn2020, mann2025squeezingclassicalantiferromagnetsquantum, Hosseinabadi2026, winter2026extensivemixedstateentanglementkinetically}.

Fiber integration provides a route toward multiple photonic interfaces within a single atom-array platform. Owing to the microscopic footprint of the cavity, multiple fiber Fabry-Pérot resonators could be incorporated within the field of view of a single microscope objective. Such multi-cavity architectures would allow different regions of an atomic array to couple to independent optical modes, enabling parallel readout and photon-mediated links between neighboring registers \cite{Wengerowsky2018, Covey2023, Zhang2025, Sinclair2025}. Overall, the combination of strong, controllable coupling, nondestructive readout, scalable atom number, and fiber integration provides a versatile starting point for future studies in quantum optics and technological applications.

\textit{Note:} While finalizing this manuscript, we became aware of two independent related preprints on optical tweezer arrays coupled to high-cooperativity Fabry-Pérot cavities \cite{Ye2026arxiv,picot2026extendedsingleatomtweezerarrays}.

\begin{acknowledgments}
We acknowledge technical support from the Machine Shop at TU Wien and thank all bachelor and master students for their contributions to building the experimental setup. We gratefully acknowledge discussions with M.~Serbyn. We are supported by the Austrian Science Fund (FWF) START grant Y 1436-N, by the Austrian Science Fund (FWF) quantA Cluster of Excellence, by the Quantum Austria Initiative of the Austrian Research Promotion Agency (FFG), and MCSA fellowship No. 101271465 (D.~B.).
\end{acknowledgments}

\appendix
\section{Experimental details}

\subsection{Cavity characterization}
\label{app:cavity-characterization}

The cavity is built from two optical fiber mirrors (produced by Qlibri) of $\SI{180}{\micro\meter}$ and $\SI{213}{\micro\meter}$ radius of curvature, determined from the transversal mode spectrum. We obtain the cavity linewidth from spectral measurements. The probe laser is frequency stabilized by a tunable offset lock which is calibrated using a wavemeter (Toptica HighFinesse WS8-2). The cavity supports two non-degenerate linearly polarized eigenmodes split by $\Delta \omega = 2 \pi \times \SI{30(1)}{\mega \hertz}$. 
We obtain the free spectral range of the cavity by measuring the frequency difference of two lasers \SI{780}{nm} and \SI{810}{nm} tuned to be simultaneously in resonance with the cavity, and obtain $\nu_\mathrm{FSR} = \SI{1.764}{\tera \hertz}$. From this measurement we extract the cavity length as $l = \SI{85}{\micro \meter}$.  
During experiments, we stabilize the cavity in frequency using a Pound-Drever-Hall (PDH) locking scheme employing a lock laser beam at $\SI{810}{\nano \meter}$. In order to move the cavity resonance over more than a \si{\giga \hertz} we use a voltage-controlled oscillator (VCO, Analog Devices HMC6380LC4B) and a phase modulator (Exail NIR-MPX800-LN-20) to shift the frequency of the locking beam. On top of that we also modulate sidebands at a fixed frequency of $\SI{400}{\mega \hertz}$ on the beam, which are necessary for the PDH-lock. Changing the voltage supplied to the VCO allows us to move the cavity resonance during an experimental sequence within a few milliseconds.
The circulating intra-cavity locking light has a power of about $\SI{100}{nW}$.

\begin{figure*}
    \centering
    \includegraphics{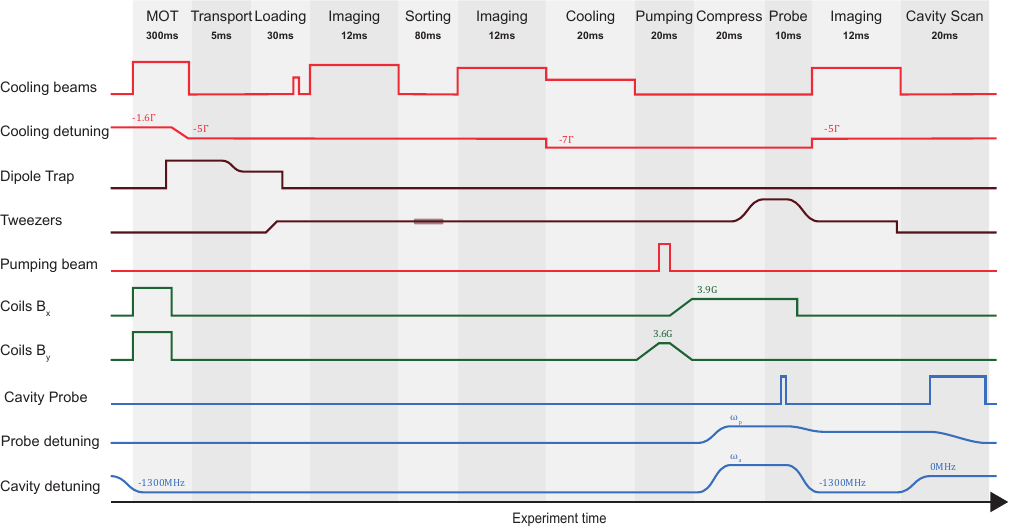}
    \caption{Experimental sequence. For the experiments with probabilistically loading multiple atoms in Figs.\,\ref{fig3} and \ref{fig5} the ``Sorting'' and ``Imaging'' afterwards are not in the sequence. The tweezer power during the probe is different for some measurements (see Sec.\,\ref{app:tweezer-characterization}) and the cavity detuning equally chosen to be resonant with the lightshifted atomic transition.}
    \label{figSequence}
\end{figure*}

\subsection{Experimental sequence}

Our experimental sequence (Fig.~\figref{figSequence}) starts by loading a three-dimensional magneto-optical trap (3D-MOT) with approximately $10^5$ $^{87}$Rb atoms in $\SI{300}{ms}$. Since the cavity length is too small to support a MOT, the 3D-MOT is about \SI{1}{\milli \meter} above the cavity center. 
During further cooling in an optical molasses stage, we load an optical dipole trap at a wavelength of $\SI{810}{\nano \meter}$ with a power of $\SI{400}{\milli\watt}$ focused to a beam waist of $\SI{20}{\micro \meter}$. This dipole trap can be moved by deflecting the beam using an acousto-optical deflector (AOD, AA DTSX-250), enabling transport of the atoms from the MOT to the cavity center within $\SI{5}{\milli \second}$. 
To trap single atoms the tweezer array is created across the cavity mode using another AOD (G\&H AODF-4150) and an arbitrary waveform generator (Spectrum M4i6631). 
We use the same laser as for the dipole trap to create the tweezer beams with a waist of $\SI{0.79}{\micro\meter}$, focussed though an objective with NA=0.5 (Mitutoyo	G Plan Apo 50X t5). 
For site-resolved fluorescence imaging two red-detuned beams with a waist of $\SI{33}{\micro \meter}$ and perpendicular linear polarization are focused through the cavity. 
We collect the scattered photons with the same microscope objective and image it on a low noise qCMOS camera (ORCA-Quest C15550-20UP). 
For loading and fluorescence imaging the cavity frequency is red-detuned by $\SI{1.3}{\giga \hertz}$ to suppress Purcell-enhanced scattering into the cavity mode.
To prepare a single atom deterministically, we reprogram the tweezer traps conditioned on a first fluorescence image.
Before measuring the cavity spectra, we define the quantization axis with a magnetic field $B_y=\SI{3.6}{G}$ perpendicular to the optical cavity and optically pump the atoms into the state $\ket{1}$ using $\sigma^+$-polarized light.
For transmission measurements we move the cavity on resonance with the atoms and quadruple the optical tweezer power for better localization of the atoms. 
We proceed to probe the cavity for $\tau = \SI{40}{\micro \second}$ and record the trace of the SPCM (Excelitas AQRH-16-FC). During probing, we apply a homogeneous magnetic field of $B_x=\SI{3.9}{G}$ along the cavity axis to define the quantization axis. 
The mean intra-cavity photon field of the probe is about $10^{-3}$ photons, which is on the order of the critical photon number. 
At the end we take a second fluorescence image of the atoms to ensure no atom loss occurred and record a trace of the empty cavity resonance to ensure it is still locked. 

\subsection{Imaging}
\label{app:imaging}
For single-atom resolved imaging the detuning of the imaging light is $-5\Gamma$ from the bare atomic resonance, where $\Gamma = 2 \gamma$ is the atomic population decay rate, and the intensity per beam is $\SI{12}{mW \per cm^2}$. Additionally, we use a repump beam (from $\ket{F=1} \rightarrow \ket{F'=2}$) with approximately half the intensity.  We eliminate unwanted scattering light on the pictures by subtracting a bright background image from all data images. 
To calculate the imaging fidelity and measure the survival rate we take two consecutive images \cite{Holman2025}. At an imaging time of $\SI{12}{ms}$, we reach a single atom detection fidelity of $99.91(3)\%$ on average for an array size of 9 tweezers in the cavity. 
The imaging fidelity across the whole  array varies by about $0.1\%$, even though the total fluorescence photons are reduced towards the edges of the array due to the small size of the imaging beams. This effect is also visible on the averaged and single-shot images in Fig.~\figref{fig1}[c], the histogram there is shown for a single site. In Fig.~\figref{fig1}[c] the colorbar for averaged image has about half of the photon counts as the single-shot image because of the tweezer loading probability of $56\%$.
The imaging survival averaged across the array is $98.0(2)\%$.

\subsection{Tweezer characterization}
\label{app:tweezer-characterization}
All characterizations here are done for the trap power used for loading and imaging of the atoms, which is then increased during probing the cavity. The trap depth is measured via the atomic lightshift of $\SI{24.3(0.6)}{\mega\hertz}$ to be $U=\SI{1.16(3)}{\milli\kelvin}$. The radial (axial) trapping frequency is determined via parametric heating by modulation of the trap power to be $\SI{94.8(8)}{\kilo \hertz}$ $(\SI{15.64(8)}{\kilo \hertz}$). From the radial trap frequency and the trap depth the tweezer's beam waist is estimated to be $\SI{790(10)}{\nano \meter}$. The temperature of the atoms after polarization-gradient cooling is measured by a release-and-recapture scheme as $\SI{34(5)}{\micro \kelvin}$. These parameters result in an average motional excitation of $\bar{n}$= 7(1), and a gaussian wavefunction spread of the atom of $\sigma_x = \SI{67(4)}{\nano \meter}$.

As shown in Fig.~\figref{fig1}[d] and Fig.~\figref{figSequence}, the power of the tweezer traps is increased within $\SI{2}{\milli \second}$ to improve the localization of the atom within the cavity mode. For tuning the cavity blockade in Fig.~\ref{fig:transmission}, the tweezer is ten times deeper compared to loading and imaging. For all other measurements, the tweezer during probing the cavity is four times deeper. The relevant parameters are summarized in the Tab.~\ref{tab:trap_parameters}.

\begin{table}[h]
    \small
    \caption{Trap parameters for different tweezer loading powers. Numbers in parentheses denote $1\sigma$ uncertainties in the last quoted digits.}
    \label{tab:trap_parameters}
    
    \begin{tabular}{cccccc}
        \toprule
        $P/P_{\rm load}$
        & Lightshift (MHz)
        & $\sigma_x$ (nm)
        & $T$ ($\mu$K)
        & $\bar{n}$
        & Usage \\
        \midrule
        1  & 24.3(6)  & 96(7)   & 34(5)    & 7(1) & -- \\
        4  & 97.0(23) & 68(5)   & 68(9)    & 7(1) & Fig.~3,4,5 \\
        10 & 243(6)   & 54(4)   & 107(15)  & 7(1) & Fig.~2 \\
        \bottomrule
    \end{tabular}

\end{table}

\noindent
The optical tweezers are aligned such that changing the RF frequency applied to the AOD translates the tweezer
along the cavity axis. We calibrate the conversion between RF frequency and
position by measuring the atom-cavity coupling at different tweezer positions over
many oscillation periods. The measured coupling is fitted to the position averaged $g(x)$ (see \ref{app:wavefunction-spread}), making use of the known spacing between adjacent
coupling maxima of $\Delta x = 390\,\mathrm{nm}$. We conclude that a change of \SI{1}{\MHz} for the AOD frequency translates the tweezers along the cavity axis by \SI{2.139(0.002)}{\micro\meter}. The positions shown in Fig.~\ref{fig_g_control} (b) are obtained by converting the respective AOD frequencies using this calibration and are plotted as displacements relative to the nearest antinode.
From the same calibration we infer the spacing between neighboring array sites in Fig.\,\ref{fig3} to be $\SI{7}{\micro \meter}$.

\subsection{Stability of maximal coupling}
\label{app:stability}

\begin{figure}
    \centering
    \includegraphics{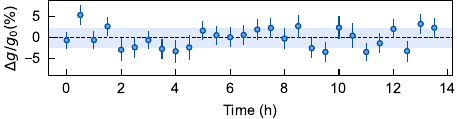}
    \caption{Stability of coupling strength. We measure the coupling strength over several hours and observe variations of $2.4\%$.}
    \label{figS_g_stability}
\end{figure}

Many cavity-QED protocols benefit from operating at maximal atom-photon coupling, which maximizes cooperativity, readout contrast, and cavity-mediated interaction rates. In a tweezer-programmable setup, this requires the maximum-coupling position to remain stable over experimentally relevant timescales. To test this stability, we position a single atom at an antinode and monitor the coupling strength over several hours using repeated transmission spectra (Fig.\,\ref{figS_g_stability}). The extracted relative coupling variation has a standard deviation of $2.4\%$. 

\section{Data analysis}
\label{app:data-analysis}

\subsection{Bayesian analysis}
\label{app:bayesian-analysis}
For all datasets, we rely on bayesian inference to provide quantitative uncertainty estimation from a limited number of collected photons~\cite{DAgostini_2003}. In this framework, we assume that the cavity transmission follows an underlying poissonian distribution and we estimate its mean $\lambda$ based on measurements. We use a conventional Gamma prior distribution for the mean photon number: $\lambda \sim \Gamma(1,0)$. Then, given that we observe $S$ photon counts in $N_\mathrm{rep}$ repetitions of the experiment, we derive the estimated posteriori distribution of $\lambda$ to be $\Gamma(1+S, N_\mathrm{rep})$ using Bayes' theorem.
Our point estimate is then this posterior mean $(S+1)/N_\mathrm{rep}$ and the equal-tailed $\SI{68}{\percent}$ credible interval yields the stated $\pm 1 \sigma$ errorbars.

To estimate an atom number discrimination fidelity between two atom numbers in Fig.~\figref{fig5}[b], we estimate the poissonian mean in this manner for each atom number and calculate the overlap of their respective distributions. We calculate the fidelity uncertainty by Monte-Carlo propagation. We  numerically sample the posteriori distribution of the poissonian means obtained from the bayesian analysis and repeat the fidelity calculation. The stated values correspond to the mean of the obtained fidelities and the uncertainties to the average of the $\SI{68}{\percent}$ confidence interval.   

\subsection{Accounting for the atomic wavefunction spread}
\label{app:wavefunction-spread}
The normalized transmission of the coupled atom-cavity system can be modeled using the expression \cite{reisererCavitybasedQuantumNetworks2015}

\begin{equation}
T(\omega_p, x) = \left|
\frac{\kappa \left(i\Delta_a + \gamma\right)}{\left(i\Delta_c + \kappa\right)\left(i\Delta_a +\gamma\right)+ g(x)^2}
\right|^2
\end{equation}
where $\Delta_{a, c} = \left(\omega_p - \omega_{a,c}\right)$ is the detuning of the probe beam from atomic (cavity) resonance and $g(x)$ is the position dependent atom-cavity coupling strength. We neglect the cavity birefringence in the data analysis. For the
theory curves in fig \ref{fig:transmission}, we assume a detuning of $\Delta_a\approx2\pi\times\SI{30}{\mega\hertz}$, comparable to $\kappa$ and $\Delta\omega$. Elsewhere $\omega_a=\omega_c$.

We account for the finite extent of the atomic wavefunction in the tweezer by assuming that the atomic position is gaussian distributed around a mean position $x_0$ with standard deviation $\sigma_x$. 

To include the effect of this spread on the transmission we use a numerical Monte Carlo approach. We sample many positions $x_i$ from this gaussian distribution and evaluate the atom-cavity coupling at each sampled position using the point-particle expression $g(x_i) = g_0 \cos(k x_i)$. We then compute the position-averaged transmission by evaluating the point-particle transmission for each sampled coupling value and taking the mean. The reported coupling strength is the root-mean-square value of the position-dependent coupling over the atomic wavepacket, $g_{\mathrm{eff}}(x_0)=\sqrt{\left\langle g(x)^2\right\rangle} $
where $g_0$ is extracted from fits of the Monte Carlo averaged transmission model to the data.

The shaded regions representing the uncertainty bounds are computed by evaluating the MC model over a grid spanning $\pm1\sigma$ interval of each parameter and taking the point-wise minimum and maximum of all resulting curves. 

\bibliography{references}

\end{document}